\documentclass[aps,prb,twocolumn,groupedaddress,showpacs,amsmath,amssymb,amsfonts]{revtex4}

\usepackage{latexsym,amssymb}
\usepackage{graphicx,color,pstricks}
\usepackage{epsfig,epsf,bm}

\definecolor{gray}{rgb}{0.7,0.7,0.7}

\begin{document}

\title{On the nature of the magnetic phase transition in a Weyl semimetal}

\author{Yu-Li Lee}
\email{yllee@cc.ncue.edu.tw}
\affiliation{Department of Physics, National Changhua University of Education, Changhua, Taiwan, R.O.C.}

\author{Yu-Wen Lee}
\email{ywlee@thu.edu.tw}
\affiliation{Department of Applied Physics, Tunghai University, Taichung, Taiwan, R.O.C.}

\date{\today}

\begin{abstract}
We investigate the nature of the magnetic phase transition induced by the short-ranged electron-electron
interactions in a Weyl semimetal by using the perturbative renormalization-group method. We find that the
critical point associated with the quantum phase transition is characterized by a Gaussian fixed point
perturbed by a dangerously irrelevant operator. Although the low-energy and long-distance physics is
governed by a free theory, the velocities of the fermionic quasiparticles and the magnetic excitations
suffer from nontrivial renormalization effects. In particular, their ratio approaches one, which indicates
an emergent Lorentz symmetry at low energies. We further investigate the stability of the fixed point in
the presence of weak disorder. We show that while the fixed point is generally stable against weak
disorder, among those disorders that are consistent with the emergent chiral symmetry of the clean system,
a moderately strong random chemical potential and/or random vector potential may induce a quantum phase
transition towards a disorder-dominated phase. We propose a global phase diagram of the Weyl semimetal in
the presence of both electron-electron interactions and disorder based on our results.
\end{abstract}


\maketitle

\section{Introduction}

Electronic materials whose low-energy excitations behave like Dirac fermions attract a lot of interests in
condensed matter physics\cite{Turner, Hosur, Vafek}. These materials are characterized by nodal points in
the Brillouin zone at which two distinct bands touch. Examples of these materials range from
graphene\cite{Castro,Katsnelson} to the surface state of a three-dimensional (3D) topological
insulators\cite{Hasan,XLQi}. These studies not only help in developing a new generation of electronic
devices, but also provide a theoretical link between condensed matter theory and high energy physics.

More recently, there has been growing interest in a close cousin of the above mentioned two-dimensional
(2D) systems --- the Weyl semimetals (WSM)\cite{Turner,Hosur,Vafek,Witczak,GangChen,Heikkila,SYXu,BQLv}.
Just like graphene, in these materials, two bands touch at certain point (the Weyl node) in the momentum
space. As usual, band touching usually leads to Dirac fermions due to the presence of the time-reversal
(T) and the space inversion (P) symmetry. To realize the Weyl fermion, which has only half a degree of
freedom of a single Dirac fermion, the involved bands must be individually non-degenerate. This requires
that either the T or the P symmetry is explicitly broken. The stability of these Weyl nodes are protected
topologically. This is closely related to the fact that each Weyl node carries a quantized monopole charge
$Q=\pm 1$ in the momentum space, which cannot be changed by a small perturbation. Moreover, since the net
``magnetic charge" must be zero in a Brillouin zone, the Weyl nodes have to appear in pairs in a
crystal\cite{Nielson}.

Since a WSM is a gapless system, it is expected that the electron-electron interaction should have
dramatic impact on its properties. Because of the vanishing density of states (DOS) at the Weyl nodes, the
short-range interaction is perturbatively irrelevant\cite{JMaciejko} while the long-range Coulomb
interaction is marginally irrelevant in the renormalization-group (RG) sense\cite{Goswami,Hosur2,Isobe}.
Therefore, to the leading order, it is reasonable to ignore the short-range interaction and the low-energy
properties of the WSM are captured by the free Weyl fermions. While for the long-range Coulomb
interaction, we only expect logarithmic corrections to the physical response functions at low frequencies
and long distances.

On the other hand, in the presence of strong electron-electron interactions, we expect a quantum phase
transition (QPT) towards some symmetry breaking phases. Previous studies based on the mean-field theory
suggest that the possible states at strong repulsive interactions include the excitonic and the
charge-density-wave (CDW) phase\cite{HWei,ZWang}. Later an unbiased pertuebative RG study on all possible
four-fermion short-range repulsions that are consistent with the symmetries of the underlying lattice was
performed\cite{JMaciejko}. By extrapolating the one-loop RG equations to the strong-coupling regime, it
was found that there is a unique direction in the asymptotic RG flow which points toward a
spin-density-wave (SDW) ground state with the characteristic momentum of the SDW state corresponding to
the momentum separation between the two Weyl nodes. Very recently, Ref. \onlinecite{LJZhai} tackled this
problem by mapping the lattice model with a strong on-site Hubbard interaction to a $t$-$J$ type model. In
terms of an appropriate mean-field treatment, it is concluded that the WSM with strong repulsions becomes
magnetic-ordered. Thus, both the perturbative RG and the strong-coupling analysis lead to a similar
conclusion.

Since the WSM phase is stable at weak coupling and an ordered phase occurs at strong coupling, the two
phases, the WSM and the ordered phase, must be separated by a quantum critical point (QCP). Up to now, it
is still unclear what is the nature of this QCP. We try to answer it in this paper. We focus on the
transition to SDW phase. We start with the minimal model of WSM which consists of two Weyl nodes with
opposite monopole charges. We propose that the critical theory is described by a single Dirac fermion
(consisting of two copies of Weyl fermions) and a complex bosonic order parameter characterizing the SDW
fluctuations. By a one-loop RG analysis, we find that the Gaussian fixed is stable at low energies so that
the correlation length exponent $\nu=1/2$. Although the QPT is characterized by a trivial fixed point in
the sense that the boson-fermion coupling is marginally irrelevant, we find that the velocities of the
fermionic quasiparticles and the magnetic excitations suffer from nontrivial renormalization effects. In
fact, their ratio approaches one in the low-energy limit. This indicates that the QCP has an emergent
Lorentz symmetry, which results in the dynamical critical exponent $z=1$.

We also investigate the stability of the above fixed point in the presence of weak disorder. The latter is
inevitable in condensed matter systems. The proposed critical theory has a chiral symmetry which is absent
in the underlying lattice model. To simplify our analysis, we consider all sorts of disorder that are
consistent with the chiral symmetry. In three dimensions, there are four types of disorder satisfying this
condition: the random chemical potential, the random vector potential, the random chiral chemical
potential, and the random chiral vector potential. We find that while the critical properties of the clean
system is stable against weak disorder for all types of the disorder mentioned above, a strong random
chemical potential or a strong random vector potential may induce a quantum phase transition toward a
disorder-dominated phase.

The disorder physics for a single non-interacting Dirac fermion is well-studied for the random chemical
potential. Already in the 1980's, Fradkin predicted the existence of a disorder-driven QPT from a
semimetallic to a diffusive metallic (DM) phase with increasing disorder strength\cite{Fradkin}. With the
renewed interest in the WSM, further theoretical studies have been performed, including the global phase
diagram of lattice models\cite{Hughes,Roy2}, the transport properties\cite{Sbierski,Syzranov}, the
calculation of critical exponents\cite{Kobayashi,Sbierski2,Pixley1,Pixley2,Syzranov2,Bera,Louvet}, and the
single-particle Green function\cite{Pixley3}. Recently, the role of the random vector potential in a
non-interacting WSM has been examined\cite{Brouwer}, and a similar disorder-driven QPT from a semimetallic
to a DM phase with increasing disorder strength is predicted.

\begin{figure}
\begin{center}
 \includegraphics[width=0.99\columnwidth]{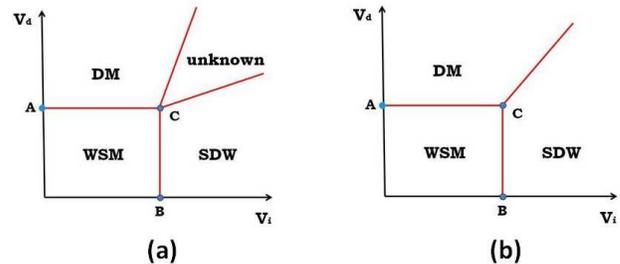}
 \caption{The schematic zero-temperature phase diagram of an interacting WSM in the presence of the random
 chemical potential or the random vector potential. $V_d$ and $V_i$ denote the disorder strength and the
 interaction strength, respectively. A denotes the QCP between the WSM and the DM phase, $B$ is the QCP
 between the WSM and the SDW phase, and C is a multicritical point. Line $\overline{AC}$ and
 $\overline{BC}$ correspond to the second-order phase transition. At strong interaction and strong
 disorder strength, there are two possibilities: (a) an unknown phase lying between the DM and SDW phase
 and (b) a direct transition between the DM and SDW phase. The shape of the phase boundaries is
 nonuniversal.}
 \label{wf1pf1}
\end{center}
\end{figure}

Since our analysis of the disorder effect is based on an effective theory of the interacting WSM, our work
complements these previous works in the sense that our results are about the disorder effect on a WSM with
{\it intermediate} strength of electron-electron interactions. Combined the knowledge on the disorder
effects of a non-interacting WSM, we propose a global phase diagram of an interacting WSM in the presence
of disorder at zero temperature as shown in Fig. \ref{wf1pf1}. At weak disorder strength and weak
interacting strength, the WSM phase is stable. By increasing the interacting strength, the WSM becomes
unstable and an ordered phase is developed. In our case, this strong-coupling phase exhibits the SDW
ordering. On the other hand, for a weakly interacting WSM, a disorder-driven QPT occurs by increasing the
disorder strength and the system turns into the DM phase at strong disorder. When both the disorder
strength and the interaction strength are strong, there may be two possibilities: an unknown phase lying
between the DM and the SDW phase (Fig. \ref{wf1pf1}(a)) or a direct transition between the DM and the SDW
phase (Fig. \ref{wf1pf1}(b)). In any case, there will be a multicritical point (point C) at which the WSM,
the DM, and the SDW phase meet with each other. The study on the nature of this multicritical point is
beyond the scope of the present work.

The rest of the paper is organized as follows. In Sec. \ref{model}, we describe the model to fix our
notation and discuss the structure of the effective theory for the QCP. The one-loop RG analysis of the
effective theory for the clean system and the effects of quenched disorder are presented in Sec. \ref{1RG}
and \ref{dis}, respectively. The last section is devoted to a conclusive discussion.

\section{The model}
\label{model}

We star with a minimal lattice model describing the non-interacting WSM whose Bloch Hamiltonian is given
by\cite{KYYang,Delplace}
\begin{equation}
 \mathcal{H}(\bm{k})=\bm{d}(\bm{k})\cdot\bm{\sigma} \ , \label{wfh1}
\end{equation}
where $\sigma_i$ with $i=1,2,3$ are Pauli matrices in the spin space and
\begin{eqnarray}
 & & d_3(\bm{k})=t_3(2+\gamma-\cos{k_1}-\cos{k_2}-\cos{k_3}) \ , \nonumber \\
 & & d_1(\bm{k})=t_1\sin{k_1} \ , ~d_2(\bm{k})=t_2\sin{k_2} \ . \label{wfh11}
\end{eqnarray}
Without loss of generality, we take $t_1,t_2,t_3>0$. $\gamma$ is a real number. One may show that when
$|\gamma|<1$, there is a pair of Weyl nodes at $\bm{k}=\pm\bm{K}$, where $\bm{K}=(0,0,k_0)$ and
$k_0=\cos^{-1}{\gamma}$. We assume that the system preserves the P symmetry but breaks the T symmetry so
that the two Weyl nodes have the same energy.

When the chemical potential coincides with the energy of the Weyl nodes, the system at low energies can be
described by the continuum Hamiltonian $H=H_0+V$, where
\begin{equation}
 H_0=\! \sum_{a=1,2,3} \! \int \! d^3x\psi^{\dagger}v_a(-i\partial_a)\alpha_a\psi \ , \label{wfh12}
\end{equation}
where $v_1=t_1$, $v_2=t_2$, $v_3=t_3\sin{k_0}$, and the Dirac matrices $\alpha_a$ are
given by
\begin{eqnarray*}
 \alpha_{1,2}=\tau_0\otimes\sigma_{1,2} \ , ~~\alpha_3=\tau_3\otimes\sigma_3 \ .
\end{eqnarray*}
Here the Pauli matrices $\tau_a$ with $a=1,2,3$ and the $2\times 2$ unit matrix $\tau_0$ describe the node
degrees of freedom. The Dirac field $\psi=[\chi_+,\chi_-]^t$ where $\chi_+$ and $\chi_-$ describe the Weyl
fermions at node $\bm{K}$ and $-\bm{K}$, respectively. $V$ consists of the short-ranged four-fermion
interactions whose actual forms can be found in Ref. \onlinecite{JMaciejko}.

In terms of the low-energy degrees of freedom, one may expand the electron operator $c(\bm{r})$ as
\begin{equation}
 c(\bm{r})\sim e^{i\bm{K}\cdot\bm{r}}\chi_+(\bm{r})+e^{-i\bm{K}\cdot\bm{r}}\chi_-(\bm{r})+\cdots \ ,
 \label{wfop1}
\end{equation}
where $\cdots$ represents the operators with scaling dimensions higher than $\chi_{\pm}$. Hence, the spin
density operator $\bm{S}=\frac{1}{2}c^{\dagger}\bm{\sigma}c$ can be written as
\begin{equation}
 S_a(\bm{r})\sim\psi^{\dagger}\tau_0\otimes\sigma_a\psi+ \! \left(e^{2i\bm{K}\cdot\bm{r}}
 \chi^{\dagger}_-\sigma_a\chi_++\mathrm{H.c.}\right) \! +\cdots \ . \label{wfop11}
\end{equation}
One may identity the operator $\chi^{\dagger}_-\sigma_a\chi_+$ as the order parameter for the SDW ordering
since a nonvanishing value of its expectation value results in
$\langle S_a\rangle=2|N_a|\cos{(2\bm{K}\cdot\bm{r}+\theta_a)}$, where
$N_a=\langle\chi^{\dagger}_-\sigma_a\chi_+\rangle$ and $\theta_a$ is the phase of $N_a$.

Based on the above observation, we propose that near the critical point lying between the WSM and the SDW
phase, the system is described by the effective theory whose Lagrangian density in the imaginary-time
formulation is of the form: $\mathcal{L}=\mathcal{L}_{\psi}+\mathcal{L}_{\phi}+\mathcal{L}_{int}$, where
\begin{eqnarray}
 \mathcal{L}_{\psi} &=& \bar{\psi}(\gamma_0\partial_{\tau}+v\gamma_a\partial_a)\psi \ , \label{wfqcl11}
 \\
 \mathcal{L}_{\phi} &=& |\partial_{\tau}\phi|^2+v^2_b|\bm{\nabla}\phi|^2+r|\phi|^2+\lambda|\phi|^4 \ ,
 \label{wfqcl12} \\
 \mathcal{L}_{int} &=& g(\phi_1\bar{\psi}\psi+i\phi_2\bar{\psi}\gamma_5\psi) \ , \label{wfqcl13}
\end{eqnarray}
with $\lambda,v_b>0$. The Dirac field $\psi$ and the complex bosonic field
$\phi=\frac{\phi_1+i\phi_2}{\sqrt{2}}$ describe the gapless fermionic quasiparticles and the SDW
fluctuations, respectively. To simplify our analysis, we take $v_1=v_2=v_3=v$. In general, $v_b\neq v$ and
$\mathcal{L}$ does not respect the Lorentz symmetry, as we would expect in condensed matter physics. The
representation of the $\gamma$-matrices is chosen to be
\begin{eqnarray}
 \gamma_0=\tau_1\otimes\sigma_3 \ , & & \gamma_1=\tau_1\otimes\sigma_2 \ , \nonumber \\
 \gamma_2=-\tau_1\otimes\sigma_1 \ , & & \gamma_3=-\tau_2\otimes\sigma_0 \ , \label{gamma1}
\end{eqnarray}
where $\sigma_0$ is the $2\times 2$ unit matrix in the spin space. One may verify that they are Hermitian
and obey the Clifford algebra
\begin{equation}
 \{\gamma_{\mu},\gamma_{\nu}\}=2\delta_{\mu\nu} \ . \label{gamma11}
\end{equation}
The matrix $\gamma_5$ is defined as $\gamma_5=\gamma_0\gamma_1\gamma_2\gamma_3=\tau_3\otimes\sigma_0$.

With the above choice of the $\gamma$-matrices, the SDW order parameter in the $z$ direction is given by
\begin{eqnarray*}
 \chi_-^{\dagger}\sigma_3\chi_++\mathrm{H.c.}=\bar{\psi}\psi \ .
\end{eqnarray*}
$\langle\bar{\psi}\psi\rangle\neq 0$ also implies dynamical chiral symmetry breaking. Hence, the SDW
ordering in the lattice model appears in the guise of chiral symmetry breaking at long distances.

By the chiral transformation
\begin{equation}
 \psi\rightarrow e^{i\theta\gamma_5/2}\psi \ , ~~\bar{\psi}\rightarrow\bar{\psi}e^{i\theta\gamma_5/2} \ ,
 \label{wfchi1}
\end{equation}
where $\theta$ is a real constant, a term involving $\bar{\psi}\gamma_5\psi$ will be generated. Hence,
$\mathcal{L}_{int}$ contains two terms whose structure is fixed by the chiral symmetry. That is,
$\mathcal{L}_{int}$ is invariant against the chiral transformation. To see this, we notice that the
operators $\bar{\psi}\psi$ and $\bar{\psi}\gamma_5\psi$ transform as
\begin{equation}
 \left(\begin{array}{c}
 \bar{\psi}\psi \\
 i\bar{\psi}\gamma_5\psi
 \end{array}\right) \! \rightarrow \! \left(\begin{array}{cc}
 \cos{\theta} & \sin{\theta} \\
 -\sin{\theta} & \cos{\theta}
 \end{array}\right) \! \! \left(\begin{array}{c}
 \bar{\psi}\psi \\
 i\bar{\psi}\gamma_5\psi
 \end{array}\right) , \label{wfchi11}
\end{equation}
under the chiral transformation [Eq. (\ref{wfchi1})]. We further write $\mathcal{L}_{int}$ as
\begin{eqnarray*}
 \mathcal{L}_{int}=g(\phi_1,\phi_2)^t \! \left(\begin{array}{c}
 \bar{\psi}\psi \\
 i\bar{\psi}\gamma_5\psi
 \end{array}\right) .
\end{eqnarray*}
Then, the invariance of $\mathcal{L}_{int}$ under the chiral transformation, Eq. (\ref{wfchi1}), requires
that $\phi_1$ and $\phi_2$ transform as
\begin{equation}
 \left(\begin{array}{c}
 \phi_1 \\
 \phi_2
 \end{array}\right) \! \rightarrow \! \left(\begin{array}{cc}
 \cos{\theta} & \sin{\theta} \\
 -\sin{\theta} & \cos{\theta}
 \end{array}\right) \! \! \left(\begin{array}{c}
 \phi_1 \\
 \phi_2
 \end{array}\right) . \label{wfchi12}
\end{equation}
Equation (\ref{wfchi12}) implies that $\phi$ transforms as
\begin{equation}
 \phi\rightarrow e^{-i\theta}\phi \ . \label{wfchi13}
\end{equation}
That is, the U($1$) transformation of the $\phi$ field corresponds to the chiral transformation, not the
charge U($1$) transformation. This chiral symmetry also determines the structure of $\mathcal{L}_{\phi}$.
We have to emphasize that this chiral symmetry is an emergent one since it is not the symmetry of the
microscopic lattice model.

The phase diagram of $\mathcal{L}$ can be seen as follows. For $r>0$, the $\phi$ field is gapped and we
may integrate it out. The resulting theory is a single species of Dirac fermions with short-range
spin-dependent four-fermion interactions. Since the four-fermion interactions are irrelevant at weak
coupling, this corresponds to the WSM phase. On the other hand, for $r<0$, $\langle\phi\rangle$ is pinned
at some nonzero value and the chiral symmetry is broken. This gives a mass to the Dirac fermion, and the
resulting phase exhibits the SDW ordering. Hence, $r=0$ corresponds to the QCP. A RG analysis is warranted
to study its critical properties.

\section{One-loop RG analysis}
\label{1RG}

In the analysis of quantum phase transitions involving gapless fermions, people usually employ the
Hertz-Millis-Moriya theory\cite{Hertz}, where the fermions are integrated out to obtain an effective
action of the order parameter, with the assumption that each term in the resulting action can be written
as a local functional of the order parameter. Such an approach, however, is shown to be incomplete due to
either the breakdown of Fermi-liquid theory\cite{Lohn}, or an infinite number of local marginal operators
being generated\cite{Abanov}. Moreover, a recent large-scale quantum Monte-Carlo study reported results
not consistent with the Hertz-Millis-Moriya theory\cite{Schattner}. Here we shall treat the fermionic and
bosonic fields on equal footing.

Under the scaling transformation $x_a\rightarrow s^{-1}x_a$ ($a=1,2,3$) and $\tau\rightarrow s^{-z}\tau$,
the scaling dimensions of various fields and parameters at the tree level are given by
$[\psi]=\frac{3}{2}$, $[\phi]=\frac{3-z}{2}$, $[v]=z-1=[v_b]$, $[r]=2z$, $[\lambda]=3(z-1)$, and
$[g]=\frac{3}{2}(z-1)$. If we choose $v$ to be RG invariant, then $z=1$. Thus, we get
$[\psi]=\frac{3}{2}$, $[\phi]=1$, $[r]=2$, and $[\lambda]=0=[g]$. We see that $r$ is a relevant
perturbation around the Gaussian fixed point (characterized by $(r,\lambda,g)=(0,0,0)$), whereas both
$\lambda$ and $g$ are marginal perturbations at the tree level.

To determine the fate of the boson-fermion coupling $g$, we calculate the RG equations to the one-loop
order. To proceed, we assume that there are $N$ species of Dirac fermions, i.e.,
$\psi\rightarrow\psi_{\alpha}$ with $\alpha=1,2,\cdots,N$, and rescale the coupling constant $g$ by
$g\rightarrow g/\sqrt{N}$. To calculate the RG equations, we separate the fields $\psi_{\alpha}$ and
$\phi_i$ ($i=1,2$) into the slow and fast modes: $\psi_{\alpha}=\psi_{\alpha<}+\psi_{\alpha>}$ and
$\phi_i=\phi_{i<}+\phi_{i>}$, where the fast modes $\psi_{\alpha>}$ and $\phi_{i>}$ consist of the Fourier
components with $e^{-l}\Lambda<|\bm{k}|<\Lambda$, while the slow modes $\psi_{\alpha<}$ and $\phi_{i<}$
consist of the Fourier components with $|\bm{k}|<e^{-l}\Lambda$. Here $\Lambda$ is the UV cutoff in
momenta and the scaling parameter $l>0$.

\begin{figure}
\begin{center}
 \includegraphics[width=0.9\columnwidth]{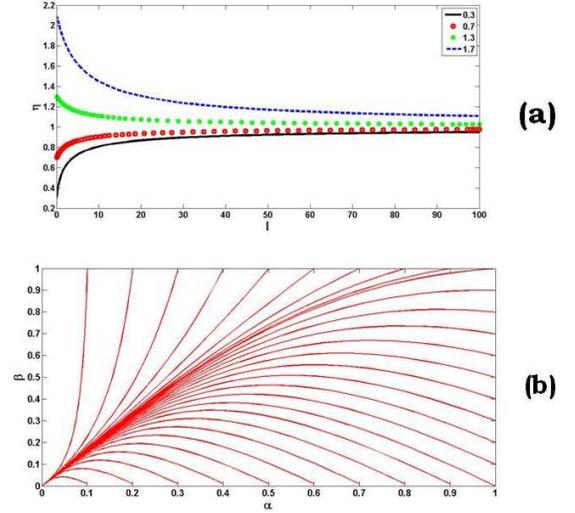}
 \caption{(a) The RG flow of $\eta(l)=v_b/v$ for various various values of $\eta(0)$ with $N=1$ and
 $\alpha(0)=0.1$. (b) The RG flow of $\alpha$ and $\beta$ with $N=1$ by setting $\eta=1$.}
 \label{wf1pf11}
\end{center}
\end{figure}

By integrating out the fast modes to the one-loop order and rescaling the the space, time, and fields by
$x_a\rightarrow e^lx_a$, $\tau\rightarrow e^{zl}\tau$,
$\psi_{\alpha<}\rightarrow Z^{-1/2}_{\psi}\psi_{\alpha}$, and
$\phi_{i<}\rightarrow Z^{-1/2}_{\phi}\phi_{i}$, we obtain the one-loop RG equations:
\begin{eqnarray}
 \frac{d\ln{v}}{dl} &=& z-1-\frac{8v^2(v-v_b)\alpha}{3Nv_b(v+v_b)^2} \ , \label{wfrge1} \\
 \frac{d\ln{v_b}}{dl} &=& z-1+\! \left(\frac{v^2}{v_b^2}-1\right) \! \alpha \ , \label{wfrge11} \\
 \frac{d\beta}{dl} &=& 3(z-1)\beta-4\beta\alpha-\frac{5v^3\beta^2}{v_b^3}+\frac{8\alpha^2}{N} \ ,
 \label{wfrge12} \\
 \frac{d\alpha}{dl} &=& 3(z-1)\alpha-2 \! \left[1+\frac{4v^3}{Nv_b(v+v_b)^2}\right] \! \alpha^2 ,
 \label{wfrge13} \\
 \frac{d\tilde{r}}{dl} &=& 2(z-\alpha)\tilde{r}+\frac{4v\beta}{\sqrt{v_b^2+v^2\tilde{r}}}-8\alpha \ ,
 \label{wfrge14}
\end{eqnarray}
where the dimensionless couplings are defined as $\alpha=\frac{g^2}{8\pi^2v^3}$,
$\beta=\frac{\lambda}{4\pi^2v^3}$, and $\tilde{r}=\frac{r}{v^2\Lambda^2}$. The wavefunction
renormalization constants $Z_{\psi}$ and $Z_{\phi}$ are chosen such that the terms
$\bar{\psi}\gamma_0\partial_{\tau}\psi$ and $|\partial_{\tau}\phi|^2$ in $\mathcal{L}$ are RG invariant.
If we take $v$ to be a RG invariant, then Eq. (\ref{wfrge1}) gives
$z=1+\frac{8(1-\eta)\alpha}{3N\eta(1+\eta)^2}$ and the other RG equations become
\begin{eqnarray}
 \frac{d\eta}{dl} &=& \! \left(\frac{1-\eta^2}{\eta^2}\right) \! \! \left[1+\frac{8\eta}{3N(1+\eta)^3}
 \right] \! \alpha \! , \label{wfrge2} \\
 \frac{d\beta}{dl} &=& 4 \! \left[\frac{2(1-\eta)}{N\eta(1+\eta)^2}-1\right] \! \alpha\beta
 -\frac{5\beta^2}{\eta^3}+\frac{8\alpha^2}{N} \ , \label{wfrge21} \\
 \frac{d\alpha}{dl} &=& -2\! \left[1+\frac{4}{N(1+\eta)^2}\right] \! \alpha^2 \ , \label{wfrge22} \\
 \frac{d\tilde{r}}{dl} &=& 2 \! \left[1- \! \left(1-\frac{8}{3N\eta(1+\eta)^2}\right) \! \alpha\right] \!
 \tilde{r} \nonumber \\
 & & +\frac{4\beta}{\sqrt{\eta^2+\tilde{r}}}-8\alpha \ . \label{wfrge23}
\end{eqnarray}
where $\eta=v_b/v$.

Equations (\ref{wfrge2}) -- (\ref{wfrge23}) have only one fixed point, the Gaussian fixed point,
characterized by $(\tilde{r},\alpha,\beta,\eta)=(0,0,0,1)$. The RG flows of $\alpha$, $\beta$ and $\eta$
are shown in Fig. \ref{wf1pf11}. A few comments on the critical properties are in order. (i) First of all,
the boson-fermion coupling $\alpha$ is marginally irrelevant around this fixed point. Hence, the
mean-field critical exponents are exact. In particular, the correlation length exponent $\nu$ is related
to the scaling dimension of $r$, leading to $\nu=1/2$. (ii) We notice that the velocity ratio $\eta$ will
flow to one. This indicates that the Lorentz symmetry is recovered at low energies. Since $v$ and $v_b$
are the velocities of the fermionic quasiparticles and the magnetic excitations in the SDW phase, this
fact can be examined by experiments. (iii) Due to the emergent Lorentz symmetry at low energies, the
dynamical critical exponent $z=1$ for this QCP.

\section{The effects of quenched disorder}
\label{dis}
\subsection{The random potential}

Now we would like to study the effects of quenched disorder on this QCP. In order to describe the disorder
effects, the Dirac fields are coupled to a random field $A(\bm{r})$ through the Hamiltonian
\begin{equation}
 H_{dis}=-v_{\Gamma} \! \int \!d^3x\psi^{\dagger}\Gamma\psi A(\bm{r}) \ , \label{wfqcl14}
\end{equation}
where $v_{\Gamma}$ measures the strength of the single-impurity potential. Since the chiral symmetry plays
an important role on the critical properties of the clean system, we would like to respect this symmetry.
In three dimensions ($3$D), there are $16$ linearly independent choices of $\Gamma$: $I$, $\gamma_{\mu}$,
$i\gamma_{\mu}\gamma_5$, and $\gamma_5$, where $I$ is the $4\times 4$ unit matrix,
$\sigma_{\mu\nu}=\frac{i}{2}[\gamma_{\mu},\gamma_{\nu}]$, and $\mu,\nu=0,1,2,3$. Among these vertices,
only half of them, i.e., $\Gamma=I,\sigma_{\mu\nu},\gamma_5$, preserve the chiral symmetry. We will focus
on these types of random potentials. We will see later that these random potentials are divided into four
classes: $\Gamma=I$, $\Gamma=\gamma_5$, $\Gamma=\sigma_{ab}$, and $\Gamma=i\gamma_0\bm{\gamma}$. The
random field $A(\bm{r})$ is nonuniform and random in space, but constant in time. Thus, it mixes up the
momenta but not the frequencies. We further assume that it is a quenched, Gaussian white-noise field with
the correlation functions:
\begin{equation}
 \langle A(\bm{r})\rangle=0 \ , ~~
 \langle A(\bm{r}_1)A(\bm{r}_2)\rangle=\Delta\delta(\bm{r}_1-\bm{r}_2) \ , \label{wfdis1}
\end{equation}
for $\Gamma=I,\gamma_5,\sigma_{ab}$, and
\begin{equation}
 \langle A_a(\bm{r})\rangle=0 \ , ~~
 \langle A_a(\bm{r}_1)A_b(\bm{r}_2)\rangle=\frac{1}{3}\Delta\delta_{ab}\delta(\bm{r}_1-\bm{r}_2) \ ,
 \label{wfdis11}
\end{equation}
for $\Gamma=i\gamma_0\bm{\gamma}$, where $a,b=1,2,3$ and the variance $\Delta $ is chosen to be
dimensionless. Let us understand the significance of these random potentials in terms of the original
lattice fermions.

First of all, the matrix $\Gamma=i\gamma_0\bm{\gamma}$ corresponds to the random vector potential, which
describes the randomness in the phases of the hopping amplitudes for the lattice fermions in the presence
of a random magnetic field. Its coupling is uniquely fixed by the gauge principle. This term breaks the T
symmetry in a fixed sample, but not on the average. This is not a problem since we have assumed the
breaking of the T symmetry in the underlying lattice model.

Next, consider the case with $\Gamma=I$, which corresponds to the random chemical potential. Since this
term couples to the two Weyl fermions equally, it describes a smooth non-staggered potential that varies
very little over each unit cell of the original lattice. This term preserves the T symmetry, but violates
the particle-hole symmetry in a fixed sample.

Let us turn into the case with $\Gamma=\gamma_5$. Clearly, this term couples to the two Weyl fermions
with opposite signs, which is thus dubbed as the random chiral chemical potential. It arises from the
staggerd component of a potential, and results in chirality imbalance in a fixed sample, but not on
average.

Finally, we consider the case with $\Gamma=\sigma_{ab}$. We notice that
$\sigma_{ab}=-\epsilon_{abc}(i\gamma_0\gamma_c)\gamma_5$, which form the three components of a chiral
vector potential. Thus, this term is dubbed as the random chiral vector potential. This term also breaks
the T symmetry in a fixed sample, but not on average. However, the three components have different
physical origins. In terms of the original lattice fermions, the term
$\Gamma=\sigma_{12}=\tau_0\otimes\sigma_3$ describes a smooth non-staggered Zeemann coupling that varies
very little over each unit cell of the original lattice, which arises from a random magnetic field in the
$z$ direction. On the other hand, the rest two components, $\sigma_{32}=\tau_3\otimes\sigma_1$ and
$\sigma_{13}=\tau_3\otimes\sigma_2$ involve chirality imbalance and Zeeman coupling simultaneously. The
latter arises from a random magnetic field in the $xy$-plane. Due to this reason, we will study the
effects of the three components separately. In fact, it suffices to consider the term
$\Gamma=\sigma_{12}$.

\subsection{The one-loop RG analysis}

Following the method employed in Ref. \onlinecite{Stauber,Huh,Wang}, the RG equations in the presence of
weak disorder can be obtained by integrating out the fast modes for the fields $\psi$, $\phi$ and then
performing the rescaling for space, time, and fields: $x_a\rightarrow e^{l}x_a$,
$\tau\rightarrow e^{zl}\tau$, $\psi_<\rightarrow Z^{-1/2}_{\psi}\psi$,
$\phi_<\rightarrow Z_{\phi}^{-1/2}\phi$, and $A\rightarrow e^{-3l/2}A$ to bring the terms
$\bar{\psi}_<\gamma_0\partial_{\tau}\psi_<$ and $|\partial_{\tau}\phi_<|^2$ back to the original forms and
to take $\Delta$ to be a RG invariant. To the one-loop order, we find that
\begin{eqnarray}
 \frac{d\ln{v}}{dl} &=& z-1-\frac{8v^2(v-v_b)\alpha}{3Nv_b(v+v_b)^2}-\gamma \ , \label{wfrge3} \\
 \frac{d\ln{v_b}}{dl} &=& z-1+\! \left(\frac{v^2}{v_b^2}-1\right) \! \alpha \ , \label{wfrge31}
\end{eqnarray}
and
\begin{eqnarray}
 \frac{d\alpha}{dl} &=& 3(z-1)\alpha-2 \! \left[1+\frac{4v^3}{Nv_b(v+v_b)^2}\right] \! \alpha^2 \nonumber
 \\
 & & -2(1+\eta_{\Gamma})\alpha\gamma \ , \label{wfrge32} \\
 \frac{d\gamma}{dl} &=& 2(z-2)\gamma+\frac{8\alpha\gamma}{N} \! \left[\xi_{1\Gamma}-\frac{v^3}
 {v_b(v+v_b)^2}\right] \nonumber \\
 & & +2(\xi_{2\Gamma}-1)\gamma^2 \ , \label{wfrge33} \\
 \frac{d\beta}{dl} &=& 3(z-1)\beta-4\alpha\beta+\frac{8\alpha^2}{N}-\frac{5v^3\beta^2}{v_b^3} \ ,
 \label{wfrge34} \\
 \frac{d\tilde{r}}{dl} &=& 2(z-\alpha)\tilde{r}+\frac{4\beta}{\sqrt{(v_b/v)^2+\tilde{r}}}-8\alpha \ ,
 \label{wfrge35}
\end{eqnarray}
where the definitions of $\alpha$, $\beta$, and $\tilde{r}$ are the same as before,
$\gamma=\Delta\tilde{v}_{\Gamma}^2$, and
\begin{eqnarray*}
 \eta_{\Gamma} &=& \! \left\{\begin{array}{cc}
 1 & \Gamma=I,\sigma_{12} \\
 -1 & \Gamma=i\gamma_0\bm{\gamma},\gamma_5
 \end{array}\right. , \\
 \xi_{1\Gamma} &=& \! \left\{\begin{array}{cc}
 \frac{v^3}{v_b(v+v_b)^2} & \Gamma=I \\
 -\frac{v^3}{v_b(v+v_b)^2} & \Gamma=\gamma_5 \\
 \frac{v^2(v+2v_b)}{3v_b(v+v_b)^2} & \Gamma=i\gamma_0\bm{\gamma} \\
 -\frac{v^2(v+2v_b)}{3v_b(v+v_b)^2} & \Gamma=\sigma_{12}
 \end{array}\right. , \\
 \xi_{2\Gamma} &=& \! \left\{\begin{array}{cc}
 1 & \Gamma=I,\gamma_5 \\
 1/9 & \Gamma=i\gamma_0\bm{\gamma} \\
 -1/3 & \Gamma=\sigma_{12}
 \end{array}\right. .
\end{eqnarray*}
If we take $v$ to be a RG invariant, then we have $z=1+\frac{8\alpha(1-\eta)}{3N\eta(1+\eta)^2}+\gamma$ and the one-loop RG equations become
\begin{eqnarray}
 \frac{d\ln{\eta}}{dl} &=& \! \left[\frac{1-\eta^2}{\eta^2}+\frac{8(1-\eta)}{3N\eta(1+\eta)^2}\right] \!
 \alpha+\gamma , \label{wfrge4} \\
 \frac{d\alpha}{dl} &=& -2 \! \left[1 \! + \! \frac{4}{N(1+\eta)^2}\right] \! \alpha^2+(1-2\eta_{\Gamma})
 \alpha\gamma , \label{wfrge41} \\
 \frac{d\gamma}{dl} &=& -2\gamma+\frac{8\alpha\gamma}{N} \! \left[\xi_{1\Gamma}-\frac{1+2\eta}
 {3\eta(1+\eta)^2}\right] \! +2\xi_{2\Gamma}\gamma^2 , ~~~~\label{wfrge42} \\
 \frac{d\beta}{dl} &=& 2 \! \left[\frac{2(1-\eta)}{N\eta(1+\eta)^2}-1\right] \! \alpha\beta
 +\frac{8\alpha^2}{N}-\frac{5\beta^2}{\eta^3} \nonumber \\
 & & +3\beta\gamma \ , \label{wfrge43} \\
 \frac{d\tilde{r}}{dl} &=& 2\tilde{r}+2 \! \left[\frac{8(1-\eta)}{3N\eta(1+\eta)^2}-1\right] \! \alpha
 \tilde{r}+2\gamma\tilde{r} \nonumber \\
 & & +\frac{4\beta}{\sqrt{\eta^2+\tilde{r}}}-8\alpha \ . \label{wfrge44}
\end{eqnarray}

\begin{figure}
\begin{center}
 \includegraphics[width=0.99\columnwidth]{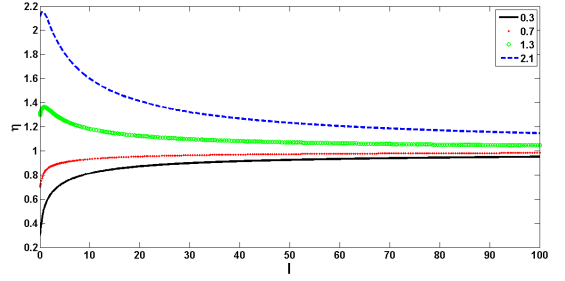}
 \caption{The RG flow of $\eta(l)=v_b/v$ for $\Gamma=I$ and various values of $\eta(0)$ with $N=1$,
 $\alpha(0)=0.1$, and $\gamma(0)=0.2$. The other types of randomness exhibit the similar behavior.}
 \label{wf1pf12}
\end{center}
\end{figure}

\begin{figure}
\begin{center}
 \includegraphics[width=0.99\columnwidth]{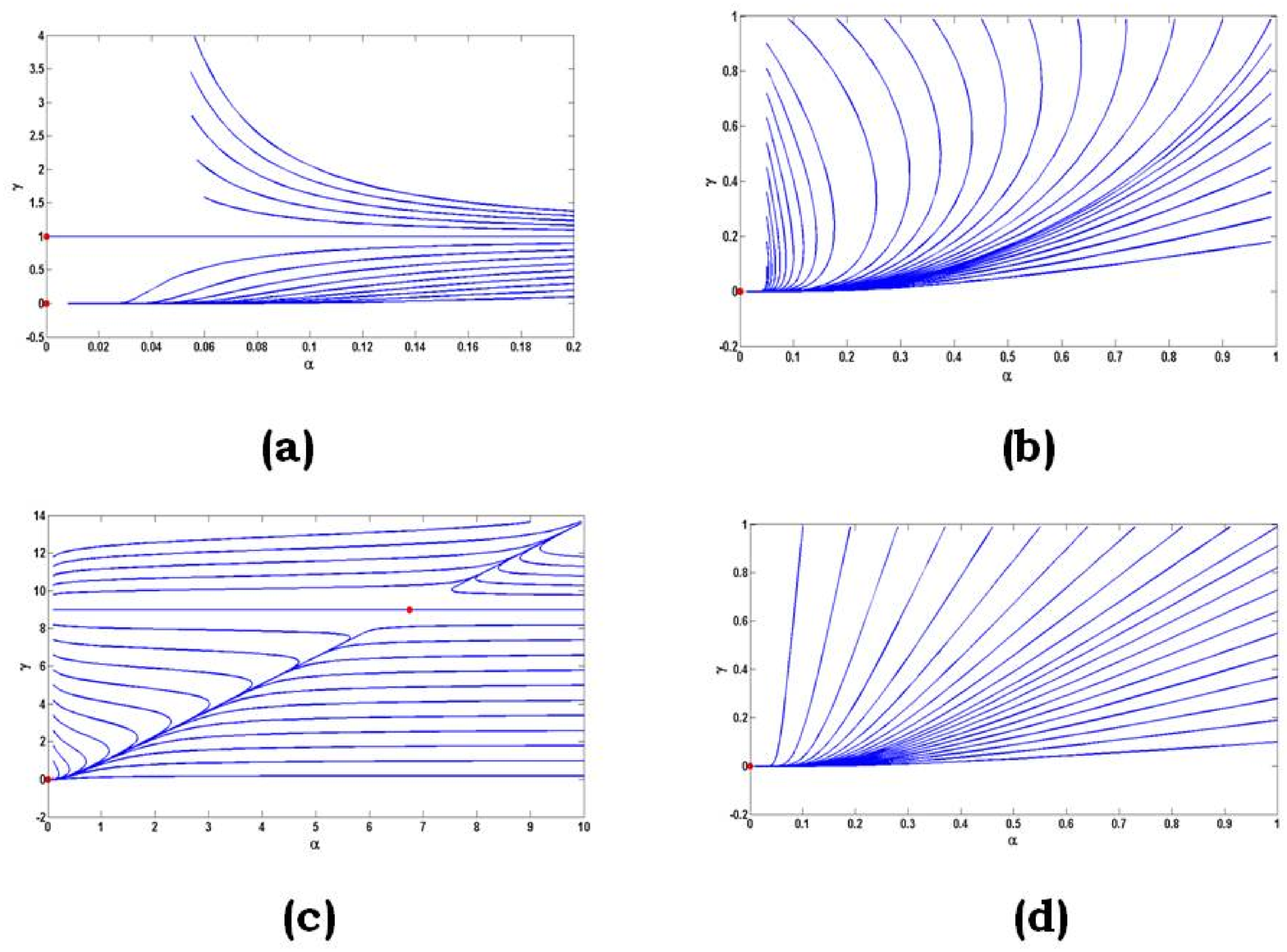}
 \caption{The RG flow of the boson-fermion coupling $\alpha$ and the disorder strength $\gamma$ for
 different types of random potentials with $N=1$ and $\eta=1$. The solid (red) circles denote the fixed
 points. (a) $\Gamma=I$, (b) $\Gamma=\gamma_5$, (c) $\Gamma=i\gamma_0\bm{\gamma}$, and (d)
 $\Gamma=\sigma_{12}$.}
 \label{wf1pf13}
\end{center}
\end{figure}

Equations (\ref{wfrge41}) -- (\ref{wfrge44}) have a fixed point characterized by
$(\tilde{r},\alpha,\beta,\gamma)=(0,0,0,0)$, the Gaussian fixed point. As we have discussed before, in the
absence of disorder, this fixed point is IR stable and the Lorentz symmetry is recovered at low energy. In
the presence of weak disorder, the stability of the Gaussian fixed point remains intact since
$\gamma$ is an irrelevant coupling around the Gaussian fixed point. Moreover, $\eta$ still flows to $1$ at
low energy, as shown in Fig. \ref{wf1pf12}. Hence, the Lorentz symmetry emerges at low energy even in the
presence of weak disorder. This results in $z=1$. The RG flows of the boson-fermion coupling $\alpha$ and
the disorder strength $\gamma$ for different types of random potentials are shown in Fig. \ref{wf1pf13}.
For clarity, we have set $\eta=1$. We discuss their behaviors in the following.

For $\Gamma=I$, the boson-fermion $\alpha$ is a marginally irrelevant coupling around the Gaussian fixed
point and will flow to zero at low energy. On the other hand, the RG equation for the disorder strength
$\gamma$ has two fixed points: $\gamma=0,1$. The former is IR stable, while the latter is IR unstable.
Hence, for weak disorder order, $\gamma$ is an irrelevant coupling, while it becomes relevant when its
value is beyond some critical one $\gamma_c$, and we get $\gamma_c=1$ by extrapolating our one-loop RG
equations to strong disorder strength. The RG flow of $\alpha$ and $\gamma$ with $N=1$ and $\eta=1$ is
shown in Fig. \ref{wf1pf13}(a). Near the non-Gaussian fixed point, ($\alpha,\gamma)=(0,1)$, the critical
line is $\gamma=1$. For weak disorder, we conclude that the critical properties of the clean system remain
intact. By increasing the disorder strength, a QPT will occur. Since $\alpha=0$ at the non-Gaussian fixed
point, we expect that this disorder-driven transition lies in the same universality class as found in the
disordered non-interacting WSM, and the WSM phase becomes the DM phase at strong disorder strength.

For $\Gamma=\gamma_5$, the RG equations have two fixed points: the Gaussian fixed point
$(\alpha,\gamma)=(0,0)$ and the non-Gaussian fixed point
$(\alpha,\gamma)=(\frac{3N}{2(N-2)},\frac{N+1}{N-2})$ when $N>2$, and have only one fixed point -- the
Gaussian one when $N=1,2$. The Gaussian fixed point is IR stable, while the non-Gaussian fixed point, if
it exists, is IR unstable. Hence, at weak disorder, the critical properties of the clean system remain
intact. The non-Gaussian fixed point describes a disorder-driven transition and the system turns into a
disorder-dominated phase at strong disorder strength when $N>2$. The RG flow of $\alpha$ and $\gamma$ with
$N=1$ and $\eta=1$ is shown in Fig. \ref{wf1pf13}(b).

For $\Gamma=i\gamma_0\bm{\gamma}$, the RG flow of $\alpha$ and $\gamma$ with $N=1$ and $\eta=1$ is shown
in Fig. \ref{wf1pf13}(c). The RG equations have two fixed points: the Gaussian fixed point
$(\alpha,\gamma)=(0,0)$ and the non-Gaussian fixed point $(\alpha,\gamma)=(\frac{27N}{2(N+1)},9)$. Similar
to the random chemical potential, the Gaussian fixed point is IR stable, while the non-Gaussian fixed
point is IR unstable. Hence, at weak disorder, the critical properties of the clean system remain intact.
The non-Gaussian fixed point describes a disorder-driven transition. At strong disorder strength, the
system turns into a disorder-dominant phase. By extrapolating our one-loop RG equations to the strong
disorder strength, the boson-fermion coupling $\alpha$ will flow to the strong-coupling regime. We expect
that the quasiparticles will acquire a gap through dynamical chiral symmetry breaking and the system would
become an insulator. However, we cannot rule out the possibility that this disorder-driven transition
belongs to the same universality class as found in Ref. \onlinecite{Brouwer} and the system is in the DM
phase at strong disorder strength.

Finally, for $\Gamma=\sigma_{12}$, the only fixed point is the Gaussian one, which is IR stable. Thus, the
critical properties of the clean system remain intact at weak disorder. The RG flow of $\alpha$ and
$\gamma$ with $N=1$ and $\eta=1$ is shown in Fig. \ref{wf1pf13}(d). It is still possible to have a
disorder-driven transition at strong disorder strength. However, this situation cannot be captured by the
one-loop RG equations.

\section{Conclusions and discussions}

In this work, we investigate the nature of the magnetic phase transition in the interacting WSM by
proposing an effective theory describing the critical region. In terms of the perturbative RG method, we
find that the critical properties are of the mean-field type and the Lorentz symmetry emerges at low
energies so that $\nu=1/2$ and $z=1$. We notice that in a recent work, the QPT to symmetry-breaking phases
for an interacting WSM has been studied from a RG analysis on the purely fermionic model\cite{Roy1}.
Interestingly, the correlation length exponent $\nu=1/2$ is also obtained, which is based on the $1/n$
expansion where $n=1$ corresponds to the WSM. We also study the effects of weak disorder on this QPT, and
show the stability of this QCP against weak disorder. The situation where the strong-coupling phase
exhibits the CDW or excitonic ordering can also be studied in a similar way by introducing the
corresponding order parameter and treating the order parameter and the fermionic quasiparticles on equal
footing.

The interplay between electron-electron interactions and disorder is not clear for the WSM. Our work sheds
some light on the global phase diagram, as shown in Fig. \ref{wf1pf1}. At weak disorder strength and weak
short-range repulsive interactions, the WSM phase is stable. By increasing the disorder strength, a QPT
from the WSM to a disorder-dominated phase may occur for the random chemical potential, the random vector
potential, or the random chiral chemical potential (for $N>2$) by extrapolating our one-loop RG equations.
For other types of disorder respecting the chiral symmetry, the WSM phase may be still stable for strong
disorder strength.

On the other hand, by increasing the interaction strength, a QPT from the WSM phase to a symmetry-breaking
phase (the SDW phase in our case) occurs. This QPT belongs to a non-Gaussian universality class within the
framework consisting only of the Weyl fermions\cite{JMaciejko,Roy1}. Nevertheless, according to our
analysis, this QPT turns into a Gaussian universality class by introducing the order-parameter
fluctuations. Furthermore, the critical properties of this QCP are immune to weak disorder.

By increasing the disorder strength along the phase boundary between the WSM and the SDW phases, we expect
the existence of a multicritical point at which the WSM, the DM, and the SDW phases meet each other. The
nature of this multicritical point is not clear at this moment. Based on the global topology of the phase
diagram, there are two possibilities when both interaction and disorder strengths are strong: (a) The
simplest scenario is that there is a direct continuous phase transition from the DM phase to the SDW
phase. (b) It is possible that there is an unknown intermediate phase existing between the DM and the SDW
phase. The nature of this intermediate phase is unclear. However, it may be insulating due to the
dynamical chiral symmetry breaking by extrapolating our one-loop RG equations. If this is indeed the case,
then the transition between this phase and the DM phase should be a continuous one. To firmly answer
whether the above speculations are true or not, one way is to find an effective field theory which
addresses the critical properties of the multicritical point directly. This is beyond the scope of the
present paper, and much work remains to be done to clarify these issues.

\acknowledgments

The works of Y.L. Lee and Y.-W. Lee are supported by the Ministry of Science and Technology, Taiwan,
under grant number MOST 105-2112-M-018-002 and MOST 105-2112-M-029-003, respectively.


\end{document}